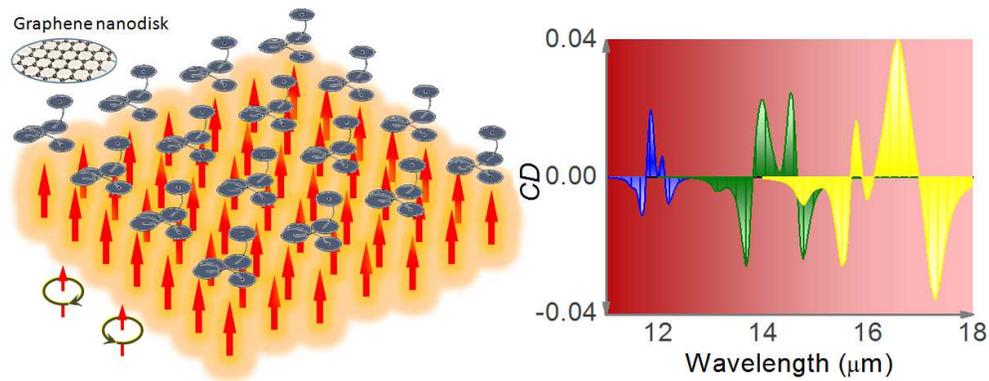

# Mid-infrared Plasmonic Circular Dichroism Generated by Graphene Nanodisk Assemblies

*Xiang-Tian Kong,[†,‡] Runbo Zhao,[†] Zhiming Wang,*[†] Alexander O. Govorov*[‡]*

[†]Institute of Fundamental and Frontier Sciences, University of Electronic Science and Technology of China, Chengdu 610054, China

[‡]Department of Physics and Astronomy, Ohio University, Athens, Ohio 45701, United States

**ABSTRACT**

It is very interesting to bring plasmonic circular dichroism spectroscopy to the mid-infrared spectral interval, and there are two reasons for this. This spectral interval is very important for thermal bio-imaging and, simultaneously, this spectral range includes vibrational lines of many chiral biomolecules. Here we demonstrate that graphene plasmons indeed offer such opportunity. In particular, we show that chiral graphene assemblies consisting of a few graphene nanodisks can generate strong circular dichroism (CD) in the mid-infrared interval. The CD signal is generated due to the plasmon-plasmon coupling between adjacent nanodisks in the specially designed chiral graphene assemblies. Because of the large dimension mismatch between the thickness of a graphene layer and the incoming light's wavelength, three-dimensional configurations with a total height of a few hundred nanometers are necessary to obtain a strong CD signal in the mid-infrared range. The mid-infrared CD strength is mainly governed by the total dimensions (total height and helix scaffold radius) of the graphene nanodisk assembly, and by the plasmon-plasmon interaction strength between its constitutive nanodisks. Both positive and negative CD bands can be observed in the graphene assembly array. The frequency interval of the plasmonic CD spectra overlaps with the vibrational modes of some important biomolecules, such as DNA and many different peptides, giving rise to the possibility of enhancing the vibrational optical activity of these molecular species by attaching them to the graphene assemblies. Simultaneously the spectral range of chiral mid-infrared plasmons in our structures appears near the typical wavelength of the human-body thermal radiation and, therefore, our chiral metastructures can be potentially utilized as optical components in thermal imaging devices.

## 1. INTRODUCTION

Strong chiroptical effects are certainly desirable for future nanotechnology and have gained broad interests in the scientific world, ranging from stereochemistry[1,2,3] to nano-optical sensing and detection[4,5,6,7] and to metamaterials.[8,9,10,11,12] The optical response of a nanoscale chiral object is intrinsically sensitive to the polarization state of the electromagnetic field, giving rise to different absorbances of left circular polarization (LCP) and right circular polarization (RCP) of light. The optical circular dichroism (CD) is ubiquitous in nature. Technically, the differential absorbance of LCP and RCP is measured by the circular dichroism spectroscopy in UV-visible-near IR spectral interval and vibrational circular dichroism spectroscopy in the mid-IR spectral interval. Optical CD of natural molecule systems (e.g., DNA, peptides) can be strong in UV, but is often weak for longer wavelengths, especially in the mid-infrared spectral interval. For example, the differential absorbance of LCP and RCP in mid-infrared for molecule systems is typically five orders smaller than their absorbance of an unpolarized light ($\Delta A \sim 10^{-5} A$).[5,13,14]

Plasmonic nanosystems of Au and Ag have been devised to enhance chiroptical effects in the visible spectral interval.[15,16,17,18,19,20,21,22] Strong plasmonic CD signals can be generated by various strategies, such as aligning plasmonic nanocrystals in three-dimensional chiral scaffolds,[23,24,25,26,27,28,29,30] building chiral plasmonic nanocrystals,[31,32,33] fabricating three-dimensional or planar chiral plasmonic nanostructures,[34,35,36,37,38,39] linking chiral molecules, such as DNA[40] and peptides,[41,42] with plasmonic nanocrystals,[43,44] and combining achiral plasmonic nanostructures with chiral molecule layers.[4,45] The enhanced CD with the plasmonic nanosystems in these strategies can be related to several physical mechanisms: (1) Coulomb and electromagnetic interaction between metal nanoparticles in chiral metal nanoparticle assemblies can create strong CD;[24,25] (2) Chiral shapes of metal nanocrystals can induce mixing of plasmonic harmonics in the metal, generating strong plasmonic CD;[31] (3) A plasmonic nanocrystal attaching to a chiral molecule can strongly enhance the CD response of the molecular component,[43,46] and the plasmon excitations can strongly interact with dipoles of chiral molecules and become chiral, creating new plasmonic lines in a CD spectrum;[44] And (4) highly twisted near fields (superchiral fields[47]) of surface plasmons can significantly enhance the optical activity of chiral molecules.[48]

In this Letter, we propose to use graphene nanostructures to generate large plasmonic CD in the mid-infrared spectrum interval. Doped graphene nanostructures can sustain intense plasmons in the mid-infrared spectrum interval.[49,50] Like surface plasmons of noble metal nanoparticles, graphene plasmons are tightly confined electromagnetic modes that can be excited by directly impinging light on the graphene nanostructures. Graphene plasmons supported by closely separated graphene nanostructures can strongly interact with each other via the near-field electromagnetic coupling.[51] Graphene nanostructures, such as nanodisks and nanoribbons, can be accurately fabricated and transferred to various substrates by today's nanotechnology.[49,50,52,53,54,55] One advantage of using graphene as the plasmonic system to generate large CD is that the plasmonic resonance of a graphene nanostructure can be tuned by chemical doping or electrical doping.[49,50,52] This allows us to tune a mid-infrared plasmonic CD spectrum mediated by graphene nanostructures.[56]

To generate strong CD effect in mid-infrared, we build a chiral unit by placing graphene nanodisks along a helix scaffold. Figure 1 schematically shows an array of the chiral graphene nanodisk assemblies with the name conventions of the two polarization states, LCP and RCP, of the incident mid-infrared light. The light perpendicularly illuminates the assembly array. The total thickness of the assembly array is several hundred nanometers, which is much less than the wavelength of the incident light. Here all the nanodisks in the assembly array are with an equal diameter. The nanodisks are evenly spaced along the light propagation direction, and also along the azimuthal direction of the helix scaffold. The array period is large compared with the helix radius so that any interaction between neighboring assemblies can be negligibly small. To fabricate the chiral graphene nanodisk assembly array, one can repeatedly transfer a layer of graphene nanodisk array to the predefined position on the substrate and then deposit a layer of dielectric as the new substrate. During this process, one should apply the graphene transfer techniques with high-precision control of the positions and twist angles.[53,54] Another possibility to fabricate such structures comes from the bio-assembly based, for example, on the DNA-origami approach, like it was done for gold nanoparticle DNA-assembled helices and other structures.[21,23]

The CD effect of the graphene nanodisk assemblies originates from the electromagnetic interaction between the nanodisks near the resonance wavelengths of graphene plasmons. A strong CD effect can be observed with two main conditions: (1) strong coupling between adjacent nanodisks in the assembly and (2) large geometrical scales (radius and total thickness) of the assembly. In particular, the large thickness along the light propagation direction is important for a strong differential response of the assembly to the LCP and RCP of incoming light. Theoretically, the two conditions can be simultaneously met by either increasing the number of nanodisks in the assembly, or enlarging the diameter of the nanodisks in the assembly, or both. But for practical cases, a chiral unit would only contain a few nanodisks to alleviate the fabrication difficulty. Then these two conditions have contradictory requirements on the geometrical parameters of the nanodisk assembly. The first condition requires a compact inter-nanodisk space; while the second condition requires a large inter-nanodisk space if we fix both the number of graphene nanodisks in a chiral unit and the nanodisks diameter. Hence an optimal geometrical configuration of a nanodisk assembly always exists for a practical case. Here we focus on the chiral units containing only a few nanodisks. Typically, the differential absorbance of a graphene nanodisk assembly array for LCP and RCP is on the order of $10^{-2}$ near the graphene plasmon resonance wavelengths ($\Delta A \sim 10^{-2} A$).

Potential applications of strongly chiral mid-IR nanostructures are two-fold. As it was mentioned above, chiral and non-chiral plasmonic assemblies with IR plasmons can be potentially used to reveal and enhance chiral vibrational excitations of important biomolecules. To the best of our knowledge, experimental developments in this direction were not reflected in the literature. The second potential application of chiral IR plasmons in graphene assemblies concerns processing of optical signals in devices for thermal bio-imaging and thermal motion sensing. We note that gold-based chiral metamaterials can also operate in the mid-IR spectral interval.[34,39] However optical devices designed using carbon-based nanomaterials, like we propose here, can be potentially much cheaper than optical systems based on noble metals. Other advantages of the graphene-based optics are a relatively low rate of energy dissipation and the relatively sharp plasmon features.

Currently, the mid-IR technology for human-body imaging devices grows very fast because of its obvious practical importance. Interestingly, the chiral signals in our structures can appear at the wavelength (~ 8 μm) that is very close to the wavelength of the maximum of the thermal radiation of human body, 9.5 μm. At this wavelength, thermal cameras and IR motion sensors operate. We think that our chiral metamaterials can be potentially useful as components for mid-IR optics. One interesting application of a chiral metastructure operating in the IR interval at ~ 1.4 μm (at the boundary between the near-IR and short-wavelength IR intervals) was reported in Ref. 57. The chiral detection was done using hot-electron photocurrents in a Schottky device and this photocurrent measurement was applied to read out a hidden image coded with two chiral patterns. Another related, emerging application of nanomaterials is metasolutions of plasmonic nanoscale complexes with engineered optical transmission spectra in the IR range.[58,59] Those are just two recent examples of emerging applications of chiral and non-chiral metamaterials and metasystems designed for the IR spectral interval.

## 2. THEORETICAL METHODS

Graphene, as a one-atom-layer-thick material, can be described as a surface current, given by

$$\mathbf{J}_s = \sigma_s \mathbf{E}_\parallel \tag{1}$$

where $\sigma_s$ is the surface conductivity of graphene and $\mathbf{E}_\parallel$ is the electric field parallel to the plane of graphene. The surface conductivity of graphene is given by the Kubo formula,[60] which combines the contributions of both the interband transitions of electrons and the intraband electron-photon scattering process. In mid-infrared, the interband contribution is negligibly small, and the energy absorption is almost entirely contributed by the intraband process. Furthermore, for doped graphene, the intraband surface conductivity of graphene can reduce to the Drude model[51]

$$\sigma_s = \sigma_{\text{intra}} = \frac{ie^2 E_F / \pi \hbar^2}{\omega + i/\tau}, \quad (E_F \gg k_B T) \tag{2}$$

where $E_F$ is the Fermi level of graphene measured from the Dirac point, $\tau$ is relaxation time, $k_B$ is Boltzmann constant, and $T$ is temperature. We assume a moderate doping level of $E_F = 0.4$ eV and relaxation time $\tau = 0.4$ ps for the doped graphene. Then the optical responses of the graphene nanodisk assembly arrays are numerically simulated with the COMSOL software. The absorbance of light is given by $A = 1 - T - R$, where $T$ and $R$ are transmittance and reflectance, respectively. Finally, CD is quantitatively described by the differential absorbance of LCP and RCP and can be written as

$$\text{CD} = \Delta A = A_{\text{LCP}} - A_{\text{RCP}}. \tag{3}$$

## 3. RESULTS AND DISCUSSIONS

To build a chiral unit, the graphene nanodisk assembly has to contain three or more graphene nanodisks. By closely placing the graphene nanodisks along a helix scaffold, the plasmonic oscillations with respect to individual nanodisks couple to one another, giving rise to distinct patterns of surface current density for LCP and RCP, and producing a nonzero CD effect. The electromagnetic coupling of plasmonic resonance modes in graphene nanodisks is very sensitive to their distance and relative directions. In this section, we study the differential absorbance $\Delta A$ of the graphene nanodisk assembly arrays in terms of the vertical distance between two adjacent nanodisks in an assembly, $d_v$, the number of nanodisks in an assembly, $N$, the radius of the helix scaffold, $r_{\text{helix}}$, the nanodisk diameter, $D_g$, and the refractive index of the matrix, $n$. These parameters affect both the plasmon-plasmon coupling strength between nanodisks in a chiral assembly and the overall size of the assembly. And they together determine the induced CD strength and CD spectral features.

*3.1 Effect of vertical distance between two nearest nanodisks in a chiral assembly*

Figure 2 shows the spectral features and CD effect of a series of chiral assembly arrays with each unit containing three graphene nanodisks (trimer, $N = 3$). The relative center positions of the nanodisks in a trimer unit are located at $(r, \phi, z)_p = (r_{helix}, 2\pi(p-1)/3, (p-1)d_v)$, with $p = 1, 2, 3$. We gradually increase the vertical separation ($d_v$) between adjacent nanodisks, while keeping the radius of the helix scaffold and diameter of graphene nanodisks constant ($r_{helix} = 80$ nm and $D_g = 100$ nm). By doing so, the total height of the assembly, $H = (N-1)d_v$, gradually increases, making each part of the assembly tends to interact more differently with LCP and RCP of light. However, the electromagnetic coupling between adjacent nanodisks gradually weakens.

Figure 2a shows the absorbance spectra of the assembly arrays with right circularly polarized mid-infrared incident light. The absorbance spectrum of the nanodisk monomer array, which is the graphene nanodisk array with one 100nm-nanodisk in each unit cell, is also shown for comparison. For the monomer array, the absorbance peak due to the dipolar plasmonic resonance occurs at $\lambda_0 = 8.02$ μm (dashed gray line). The trimer array with weak coupling effect (e.g., $d_v = 200$ nm) also shows the single absorption peak at this wavelength. This absorbance peak splits into two peaks when all the nanodisks strongly interact with each other in the trimer unit (e.g., when $d_v < 20$ nm). As the inter-nanodisk distance $d_v$ increases from zero, the plasmonic resonant modes associated with the individual nanodisks gradually decouple from each other and the two absorbance peaks begin to shift towards each other. The coupling strength of the uppermost and nethermost nanodisks is much weaker and decreases faster than that of the two nearest neighboring nanodisks. As a consequence of the different decoupling rates, two new absorbance features emerge as $d_v$ further increases (e.g. when $d_v \sim 40$ nm). The second and third absorbance lines, which are due to the coupling between the uppermost and nethermost nanodisks, approach each other faster than the first and fourth lines. They merge into one peak when $d_v$ increases to ~ 80 nm. Eventually when $d_v$ is large enough (e.g., $d_v = 200$ nm), all the absorbance lines merge into one peak. Furthermore, the plasmon-plasmon coupling strength can be easily shown by the distance between the splitting plasmonic absorbance peaks in the absorbance spectrum.

Figure 2b shows the surface current density distributions induced by LCP and RCP in the trimer units of the graphene nanodisk assembly arrays. The graphene nanodisk diameter determines the decay length of the plasmonic fields into the dielectric matrix. When $d_v$ is less than the decay length of graphene plasmons (e.g., $d_v$ = 80 nm), the coupling effect between the graphene plasmons supported by adjacent nanodisks is essential. And the surface currents are aligned in chiral patterns that are very distinct between the cases of LCP and RCP illuminations. In contrast, when $d_v$ is much greater than the decay length of graphene plasmons (e.g., $d_v$ = 200 nm), the surface currents do not show obvious chiral alignment with either LCP or RCP of light, indicating that the coupling effect between nanodisks becomes negligibly small. The surface current density pattern of each graphene nanodisk in the trimer resembles that of an isolated graphene nanodisk illuminated by a circularly polarized light. This demonstrates the importance of strong plasmon-plasmon interaction to the generation of large CD effects. In addition, a three-dimensional chiral frame is necessary for breaking the translational and rotational symmetries of a planar two-dimensional structure (i.e., when $d_v$ = 0) and creating CD effects.

Figure 2c shows the CD spectra with various $d_v$ values. Both positive and negative CD bands are generated by the graphene nanodisk assembly arrays near their plasmonic resonance wavelengths. As seen, both weak plasmon-plasmon interaction (e.g., $d_v$ = 200 nm) and a short height of the assemblies (e.g., $d_v$ = 10 nm, 20 nm) are detrimental to the CD effect. Figure 2d shows the maximum CD effect in terms of $d_v$. When increasing $d_v$ from 0 to 200 nm, the absolute values of both the maximum of the positive band(s), $(\Delta A)_{max}$, and the minimum of the negative band(s), $(\Delta A)_{min}$, first increase from zero, then reach a maximum, and decrease to almost zero. The optimal vertical inter-nanodisk distance lies near $d_v = 80$ nm, and the optimal CD is $\Delta A \sim 0.02$ for the graphene nanodisk trimer arrays. This non-monotonic behavior reflects the fact that the CD effect in our structures comes from the two factors: Inter-nanodisk electromagnetic coupling and geometrical chirality. For small vertical inter-unit distances, the geometry becomes non-chiral and, for a large spacing, the coupling becomes weak.

*3.2 Effect of number of graphene nanodisks in an assembly*

In general, to ensure strong plasmon-plasmon interactions, more nanodisks are needed in a chiral unit when increasing the assembly height. Therefore, the requirements of strong plasmon-plasmon interaction and large assembly height can be simultaneously satisfied by using more graphene nanodisks in an assembly unit. When increasing the number of graphene nanodisks in an assembly, the CD effect can be improved mainly due to two reasons: (1) The plasmon-plasmon coupling between two nearest nanodisks in an assembly can be strengthened as the inter-nanodisk distance decreases in both the vertical and the radical directions; And (2) the plasmonic mode associated with one nanodisk in an assembly can interact not only with its nearest neighboring nanodisk plasmons, but also with its second nearest neighbors in the assembly.

Figure 3 shows the effect of the number of graphene nanodisks in an assembly unit on the CD effect of the chiral assembly arrays ($N = 4, 5, 6, 7$). The relative coordinates of the centers of the graphene nanodisks in each chiral unit are given by $(r, \phi, z)_p = (r_{helix}, 2\pi(p-1)/(N-1), (p-1)d_v)$, with $p = 1, 2, ..., N$. The vertical inter-nanodisk distance is related with the assembly height by $d_v = H / (N-1)$. And the assembly heights are far less than the resonance wavelengths of graphene plasmons. With a fixed assembly height ($H$ = 100 nm, 200 nm, or 300 nm), the differential absorbance, $\Delta A$, near the plasmonic resonance wavelengths monotonously increases with the increasing number of nanodisks in an assembly unit, further confirming the beneficial role of the plasmon coupling effect in producing a change in absorbance. For example, when increasing $N$ from 4 to 7 in the 200nm-height graphene nanodisk assembly array (Figure 3b), the maximum of positive CD bands enhanced by 102%, from $(\Delta A)_{max}$ = 0.0216 to $(\Delta A)_{max}$ = 0.0437. And as $N$ increases, the plasmonic CD bands broaden and the CD spectral features become more complex. In particular, the CD spectral broadening effect is prominent for thin assemblies (e.g. when $H$ = 100 nm) when a graphene nanodisk in a chiral assembly can strongly interact with every other nanodisk in the assembly via the plasmon-plasmon coupling.

When $N$ is large and the inter-nanodisk space is compact enough (e.g., $N$ = 6 or 7 for $H$ < 300 nm), the plasmonic CD strength is less strongly improved by increasing the number of nanodisks in an assembly unit. Then the CD effect primarily relies on the assembly height. For example, when increasing the number of nanodisks in the chiral unit of the 100nm-height nanodisk assembly array from $N$ = 6 to $N$ = 7, the maximum of the positive CD band is only improved by 24%, from $(\Delta A)_{max}$ = 0.0184 to $(\Delta A)_{max}$ = 0.0229. In contrast, when increasing the assembly height from 100nm to 300nm, the maximum of the positive CD band can be enhanced by 134%, from $(\Delta A)_{max}$ = 0.0184 to $(\Delta A)_{max}$ = 0.0430, for $N$ = 6. Within our parameter range, the strongest CD effect with $(\Delta A)_{max}$ = 0.0606 and $(\Delta A)_{min}$ = −0.0628 at incident wavelengths of $\lambda_0$ = 7.8 μm and 8.06 μm, respectively, can be generated with $H$ = 300 nm and $N$ = 7.

*3.3 Effect of radius of helix scaffold*

Figure 4 shows the absorbance spectra and the CD spectra of graphene nanodisk assembly arrays with various radii of the helix scaffold and with a fixed number of nanodisks in a chiral assembly ($N$ = 5) and fixed assembly height ($H$ = 200 nm). The radius of the helix scaffold determines the horizontal separation of adjacent graphene nanodisks, which has an important impact on both the plasmon-plasmon interaction and the width of the assembly unit. As the helix radius increases from 0 (nanodisk stack) to a finite number, the scaffold becomes chiral, and the two absorbance peaks shift towards each other and hybrid with each other and eventually merge into a single peak when the helix radius becomes greater than the nanodisk diameter. The absorbance peaks originate from the coupling of the graphene plasmons supported by two or more neighboring nanodisks. Again, the plasmon-plasmon interaction strength can be positively correlated to the distance of the two plasmonic peaks in the absorbance spectra.

Either a helix scaffold radius greater than the nanodisk diameter (e.g., $r_{helix}$ = 120 nm) or a very small radius (e.g., $r_{helix}$ = 40 nm) is detrimental to the plasmonic CD effect. On the one hand, chiral alignment of graphene nanodisks induces different distributions of surface current between LCP and RCP of incident light. In this respect, a large radius of the helix scaffold would be preferred. On the other hand, a strong plasmon-plasmon interaction requires a large overlap of the graphene nanodisks in the horizontal directions. In this respect, the helix radius should not exceed the diameter of the nanodisks. As seen, helix radius of ~ 80 nm can satisfy both the requirements of strong horizontal plasmon-plasmon interactions and large lateral dimensions for the graphene nanodisk assembly units with nanodisks of $D_g$ = 100 nm.

*3.4 Effect of nanodisk diameter and matrix refractive index*

Figure 5 shows the absorbance spectra and the CD spectra of the chiral graphene nanodisk assembly arrays with various nanodisk diameters ($D_g$ = 100 nm, 140 nm, and 180 nm) in a matrix with the refractive index of $n$ = 1.5, which corresponds to a polymer or silica matrix. The number of nanodisks in an assembly is fixed as $N$ = 5, and the radius of the helix scaffold is $r_{helix}$ = 80 nm, the vertical distance between two nearest nanodisks is $d_v$ = 50 nm. The plasmonic resonance wavelengths can be tuned by varying the nanodisk diameters and/or the matrix refractive index. Here we show strong plasmonic CD effects can be generated by the graphene nanodisk assembly arrays at longer wavelengths in mid-IR. For example, when increasing the matrix refractive index from $n$ = 1 (Figure 4b) to $n$ = 1.5 (Figure 5b), the maximum of the positive CD bands shifts from wavelength of $\lambda_0$ = 8.04 μm to $\lambda_0$ = 11.86 μm for the graphene nanodisk assembly array with $H$ = 200 nm, $N$ = 5, $r_{helix}$ = 80 nm, and $D_g$ = 100 nm.

And when increasing the graphene nanodisk diameter, the plasmonic CD effect can be significantly enhanced while the CD bands shift to even longer wavelengths in mid-IR. By increasing the graphene nanodisk diameter, $D_g$, both the conditions of strong plasmon-plasmon interaction and the large lateral dimensions of the chiral assembly can be simultaneously satisfied. When increasing $D_g$, the enhanced plasmonic CD is mainly due to three reasons: (1) The plasmon-plasmon interaction is strengthened with the increasing overlap of the electromagnetic fields of adjacent graphene nanodisk plasmons; (2) The decay length of the graphene plasmons in the dielectric increases, resulting in a strengthened plasmon-plasmon coupling effect; And (3) the whole lateral scale of the chiral assembly unit increases. As a numerical example, when increasing $D_g$ from 100 nm to 180 nm, the maximum of the positive bands is enhanced by 100%, from $(\Delta A)_{max}$ = 0.0197 at $\lambda_0$ = 11.9 μm to $(\Delta A)_{max}$ = 0.0393 at $\lambda_0$ =

16.6 μm, and the minimum of the negative bands is enhanced by 216%, from $(\Delta A)_{min} = -0.0112$ at $\lambda_0 = 11.7$ μm to $(\Delta A)_{min} = -0.0354$ at $\lambda_0 = 17.3$ μm.

Broadband mid-IR responses are desirable in many areas. These results make it possible to achieve a broadband mid-IR CD spectrum by mixing different chiral units in an assembly array, where the nanodisks diameter $D_g$ varies from unit to unit. Another way to achieve broadband mid-IR CD effects is to incorporate differently sized nanodisks within a chiral unit. It has been recently reported that broadband mid-IR absorptions in the range of 8 μm to 14 μm can be generated from the coupling of graphene plasmons in differently sized graphene nanostructures in an array period.[61] Therefore in a similar way, we can expect that broadband mid-IR CD effects can be generated by designing different diameters for the nanodisks in a chiral assembly.

In addition, we note that the plasmonic CD properties of the chiral graphene nanodisk assembly arrays are also affected by the Fermi level of graphene. The plasmonic CD bands will shift to longer wavelengths when elevating the Fermi level of graphene. Thus the CD properties of the graphene-based chiral assemblies can be actively controlled by electrical doping or tuned by chemical doping. In particular, graphene plasmons are very sensitive to the molecules absorbed on graphene in the surface transfer chemical doping processes, which are reversible.[62] We can expect that very sensitive changing of the CD spectrum of the chiral device based on graphene nanodisk assemblies when exposing its top layer to a gas, e.g., nitric acid vapor.[62] Such properties make the chiral device an excellent candidate for environment monitoring and sensing applications.

## 4. CONCLUSION

We have numerically studied the mid-infrared CD effect generated by chiral graphene nanodisk assembly arrays. This strong CD effect appears when we arrange graphene nanodisks along a helical line. A strong CD effect requires both large vertical and horizontal scales of the chiral assembly and a strong near-field electromagnetic interaction via graphene plasmons. In our structures, the plasmonic resonance wavelengths can be tuned by the nanodisk diameter and the refractive index of the matrix material. Practically, these parameters should be determined by the spectrum interval of interest. When the nanodisk diameter is fixed, a particular vertical inter-nanodisk distance and a particular radius of helix scaffold exist for an optimized CD effect of the chiral assembly array. This study may be useful for mid-infrared molecular sensing and vibration circular dichroism spectroscopy of nanomaterials and, in general, for optical components of mid-IR sensors.

## AUTHOR INFORMATION

Corresponding Authors

*E-mails: zhmwang@gmail.com; govorov@ohio.edu

**Notes**

The authors declare no competing financial interest.

## ACKNOWLEDGMENTS

X.-T.K was supported by China Postdoctoral Science Foundation (2015M580778); A.O.G. was supported by the US Army Research Office (grant number W911NF-12-1-0407), by the

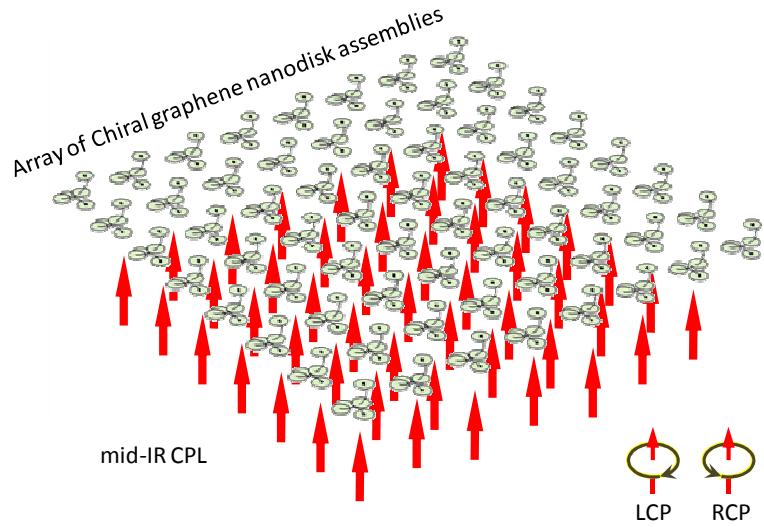

**Figure 1.** Schematic of chiral graphene nanodisk assembly array with the two circular polarization states of light, LCP and RCP. In each assembly unit, the nanodisks are placed along a helix scaffold. The array period is 400 nm in this Letter. These chiral assemblies are designed to produce strong circular dichroism in mid-infrared. The light perpendicularly illuminates the chiral array along the axis of each chiral unit. CPL: circularly polarized light.

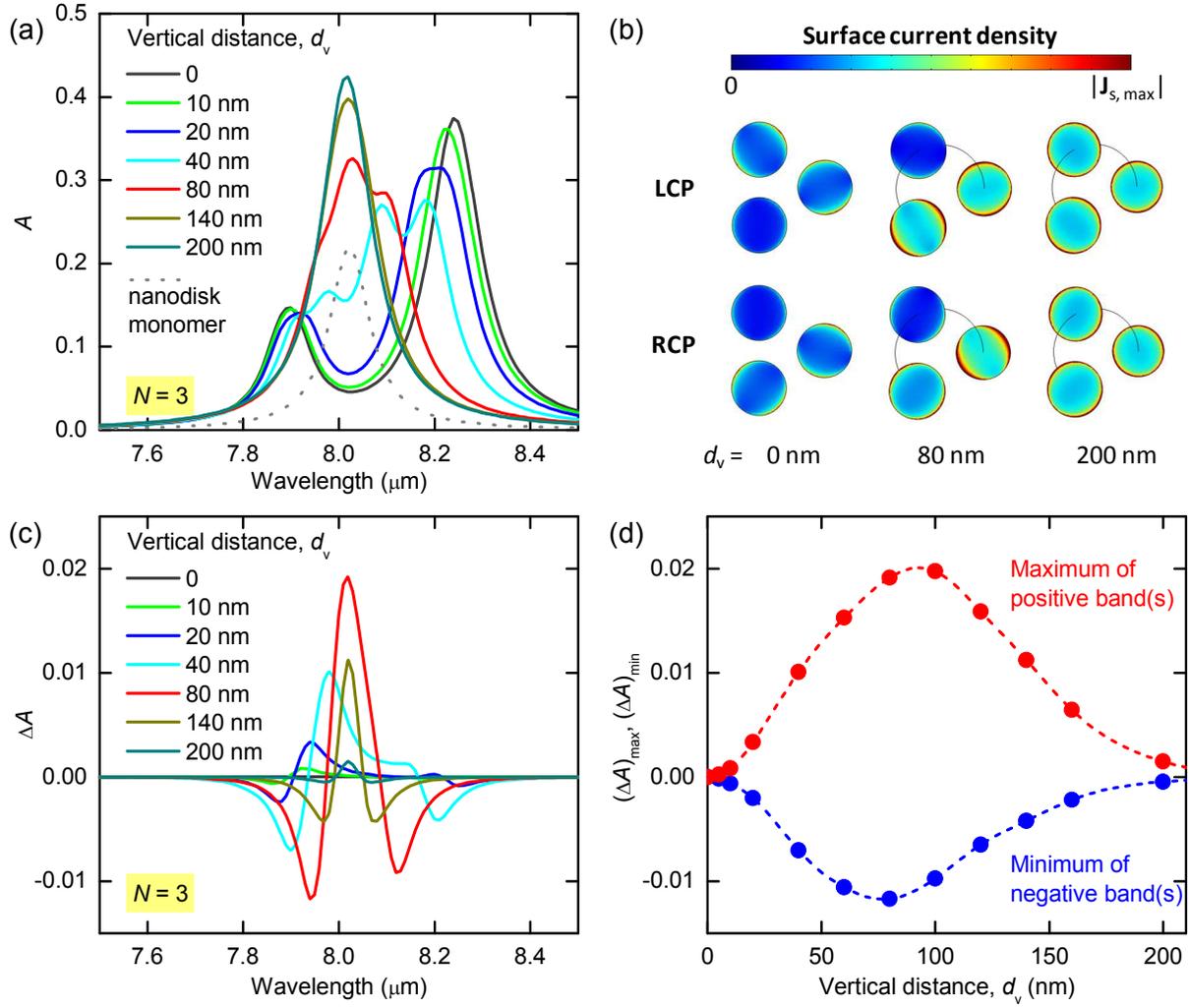

**Figure 2.** Circular dichroism generated by chiral graphene nanodisk assembly arrays with various vertical distances ($d_v$) between two nearest nanodisks in an assembly. (a) Absorbance spectra for right circular polarization, (b) Surface current density distributions (top view), and (c) Differential absorbance of the assembly arrays with various $d_v$ values. (d) Maximum and minimum of the differential absorbance with respect to the vertical distance. The number of nanodisks ($N = 3$) and radius of the helix scaffold ($r_{helix} = 80$ nm) are fixed. The diameter of graphene nanodisks is $D_g = 100$ nm. The matrix refractive index is $n = 1$.

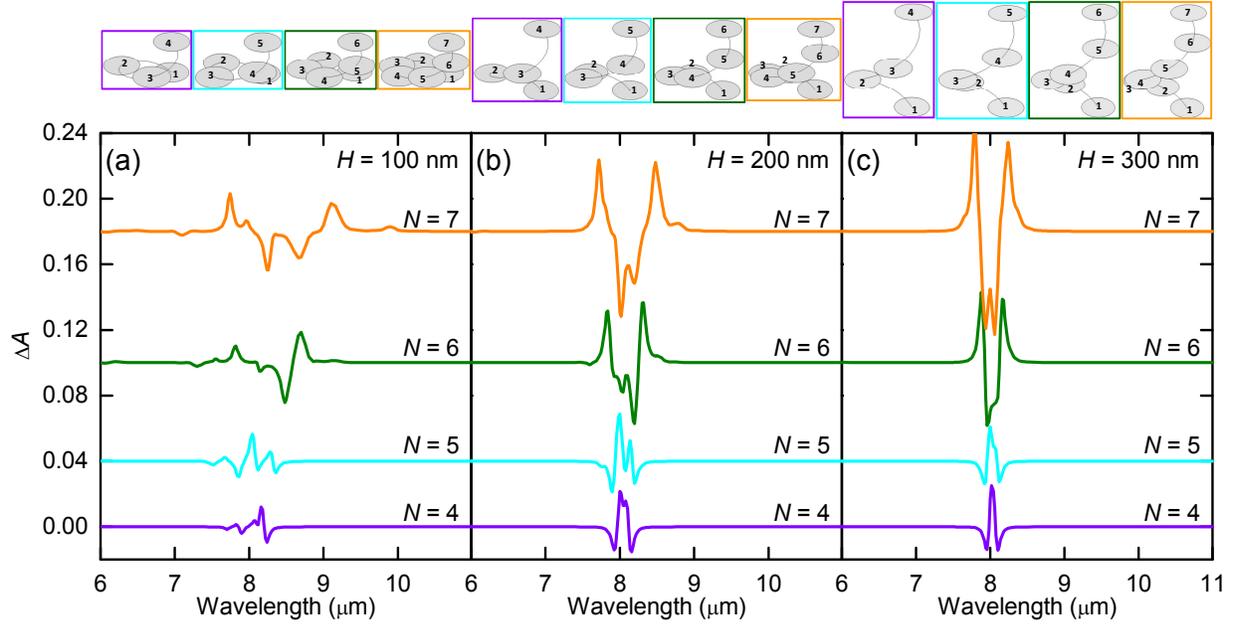

**Figure 3.** Circular dichroism of chiral graphene nanodisk assembly arrays with various numbers of graphene nanodisks ($N$ = 4, 5, 6, 7) in an assembly unit. Total height of assembly is (a) $H$ = 100 nm, (b) $H$ = 200 nm, and (c) $H$ = 300 nm. The radius of the helix scaffold is fixed at $r_{\text{helix}}$ = 80 nm. The diameter of graphene nanodisks is 100 nm. The matrix refractive index is 1. The insets show schematics of the assembly units.

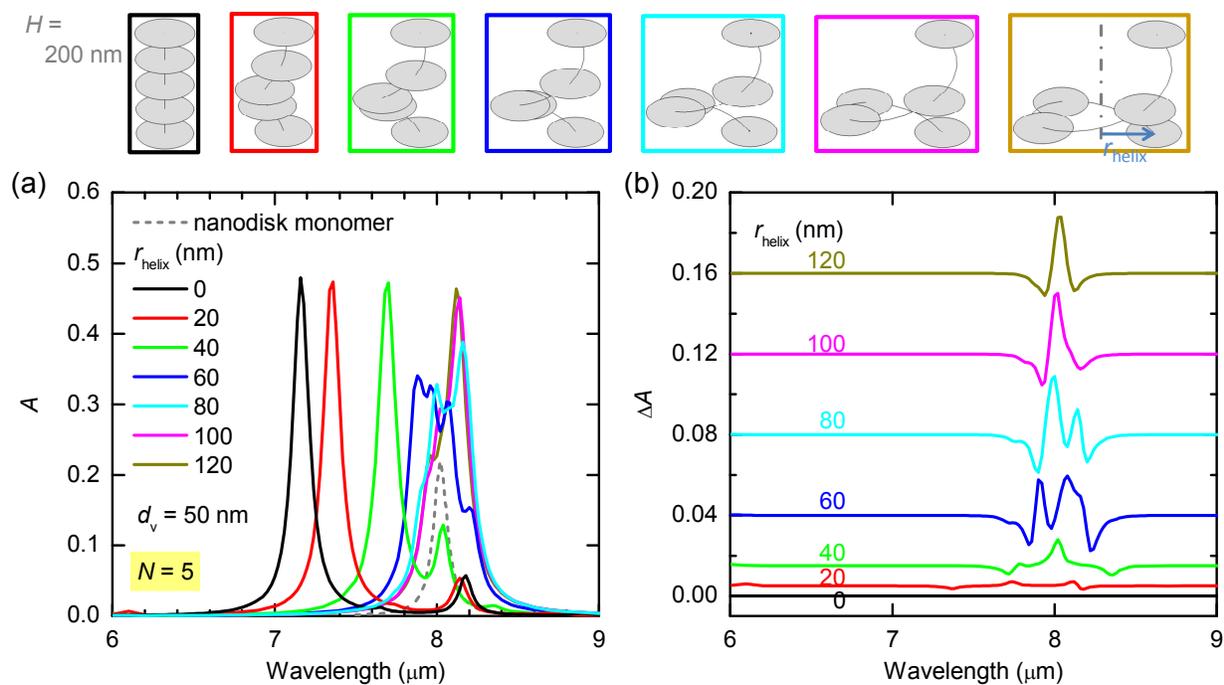

**Figure 4.** Circular dichroism of graphene nanodisk assembly arrays with various chiral helix radii. (a) Absorbance spectra for right circular polarization, and (b) CD spectra of the assembly arrays with various helix radii. The total height, the vertical distance of two adjacent nanodisks, and the number of nanodisks in an assembly are fixed. Graphene nanodisk diameter is 100 nm. Refractive index of the matrix is 1.

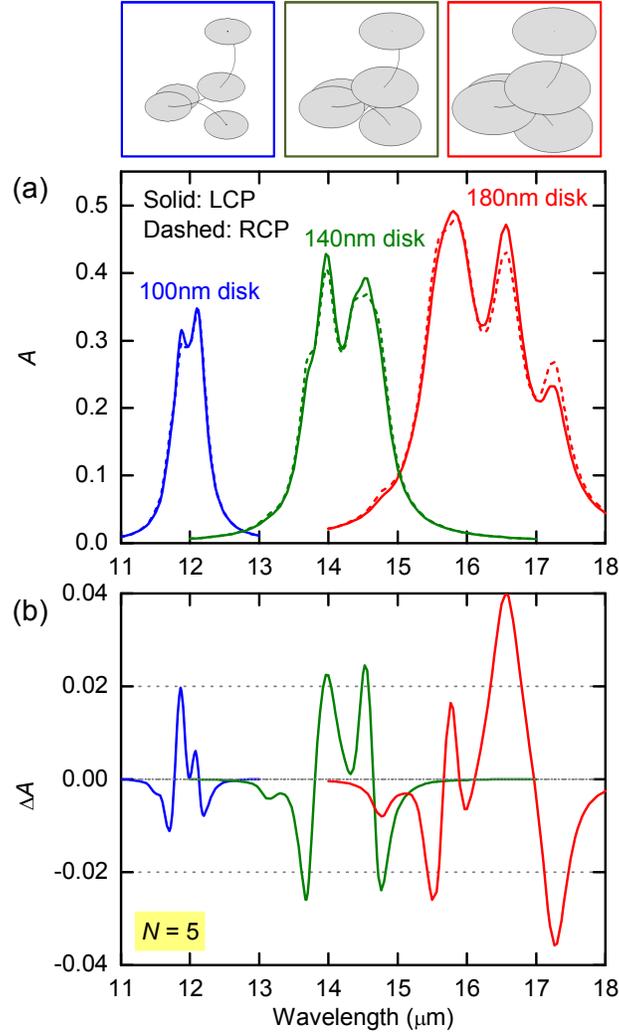

**Figure 5.** Circular dichroism of graphene nanodisk assembly arrays with various nanodisk diameters in a medium. (a) Absorbance spectra, $A_{\text{LCP}}$ (solid) and $A_{\text{RCP}}$ (dashed). (b) Differential absorbance spectra. The size of helix scaffold is fixed at $r_{\text{helix}} = 80$ nm. The number of nanodisks in an assembly is $N = 5$. The total height of the assemblies is $H = 200$ nm. The matrix refractive index is 1.5. The CD bands are tuned to wavelength intervals around $\lambda_0 \sim 12$ μm, 14 μm, and 16 μm, respectively, for graphene nanodisk diameter of $D_g = 100$ nm, 140 nm, and 180 nm.